\documentclass[10pt]{report}

\usepackage{amsmath} \usepackage{amsfonts} \usepackage{amssymb}
\begin{document}


\begin{center}

\large \textbf{An Equation For Charge Decay Valid in Both Conductors and Insulators}

\bigskip 
Albert E. Seaver
\normalsize 

7861 Somerset Ct.

Woodbury, MN 55125

E-mail: aseaver@electrstatics.us  Phone: 651-735-6760

\end{center}

\paragraph{Abstract}

Gauss' law and the equation of continuity must be satisfied 
in all materials be they solids, liquids or gases.  Most materials 
are classified as simple materials; i.e., their electrical properties 
are linear, isotropic and homogeneous.  Charge transport in these simple
 materials should be described by a constitutive equation 
known as Ohm's law.  When Ohm's law is combined with Gauss' law and the 
equation of continuity, a differential equation for volume charge density 
relaxation results.  The 
usual solution to this equation shows that charge decays exponentially 
with a relaxation time given by the material's permittivity divided by 
its electrical conductivity.   Experiments show that good conductors 
follow this exponential decay but that poor conductors (insulators) 
tend to follow a decay that initially is more hyperbolic than 
exponential.   This suggests that either Ohm's law is not valid for 
insulator materials or that a deeper understanding of Ohm's law is 
needed to explain charge decay in these less than good 
conductors.  This paper examines the latter approach and shows 
that, when all the free charges within a simple material are taken 
into account, a new closed-form equation for unipolar charge 
decay can be derived which is valid for any simple material 
be it classified as a very good conductor, or a very poor 
conductor or anywhere in between.  
For good conductors the equation reduces to the standard 
\textit{exponential law} of decay.  For very poor conductors the 
equation reduces to the Vellenga-Klinkenberg \textit{modified hyperbolic law} 
with the initial decay producing the characteristic Bustin \textit{hyperbolic law} 
of decay.  Explicit definitions for a good conductor and a good 
insulator are obtained and are used to define the range where explicit
deviations from both of these hyperbolic laws occur.
\medskip 
\begin{flushleft}
\textbf{Introduction} \end{flushleft}
Electrostatics is the study of charges including their effects and how they 
affect the local environment.  In other words, electrostatics examines 
the behavior of charges in various material media (solids, liquids and 
gases). In the absence of magnetic influences, the relevant electrical 
properties of a material are its electrical conductivity $ \sigma $ and its 
permittivity $ \epsilon $ given by $ \epsilon = \epsilon_{0}\epsilon_{r} $, where 
$  \epsilon_{0}= 8.85\times10^{-12}$ F/m is the 
permittivity of free space and $ \epsilon_{r} $ is the relative permittivity or 
dielectric constant of the material.  

When excessive charge builds up in a material, it can often have dire 
consequences such as initiation of a spark that results in a fire, 
explosion or other damage to the local environment.  For example, 
in every operation from the simple (e.g., walking) to the complicated 
(e.g., semiconductor manufacture) there is an operation time (e.g., walking 
step, ion-sputtering time, etc.) during which a material (e.g., sole of shoe, 
semiconductor chip) can become charged.  If the rate at which charge is 
deposited is greater than the rate at which charge can decay, a charge 
build-up on or within the material can occur.  As a result, it is important 
to understand charge decay in sufficient detail so as to be able to predict 
if a material constitutes an electrostatic risk in a particular operation.  

In order to gain a more fundamental insight into charge decay, this paper 
first reviews the basic concept of applying Gauss' law and the equation 
of continuity to simple materials; i.e., materials that conform to Ohm's 
law.  Then a more in-depth review of the physics which leads to Ohm's 
law is presented including the concept of electrical conductivity.  
This is followed by a more in-depth review of the partial differential 
equation that results when Ohm's law is combined with Gauss' law and 
the equation of continuity.  After defining the concept of a material 
(solid, liquid or gas; conductor or insulator) as being 
electrically neutral in the intrinsic 
state, a unipolar charge species is introduced into a simple material.  
It is shown that this charge temporarily changes the overall 
electrical conductivity, a point 
either missed or ignored in past charge decay derivations.  Once it is 
recognized that the conductivity has changed due to the charge, a closed 
form equation that describes the decay of the charge is developed.  The 
new equation reduces to the single exponential decay for a material 
classified as a good conductor.  For a good insulator the new 
equation reduces to the known "hyperbolic decay law" during the initial 
decay time.
\medskip
\begin{flushleft}
\textbf{Traditional Charge Decay Development} \end{flushleft}
Most materials can be classified as "simple" materials and this 
paper deals 
only with such materials which are defined as isotropic, linear and 
homogeneous [1].  Isotropic means that the polarization vector points 
in the direction of the electric field vector \textbf{E}. 
Linear means the electric susceptibility is a 
scalar constant so the polarization vector and electric field vector 
are linearly related, and hence, the  material's 
permittivity $ \epsilon \neq \epsilon(\textbf{E}) $ is not a 
function of the electric field.  Homogeneous means no spatial gradients 
and hence $ \nabla \epsilon = 0 $ and $ \nabla \sigma = 0 $. Within a 
material Gauss' law gives the relationship 
between the electric field \textbf{E} at a point and the net charge 
density $ \rho $ at 
that point as [1, 2]
\begin{equation} \label{e.1}
\nabla \cdot \epsilon \textbf{E} = \rho
\end{equation}
where $ \nabla $ is the usual "del" operator ([1]; pp. 39-43).  
The general continuity equation includes source terms (e.g., rate 
of charge generation) and sinks terms (e.g., rate of charge 
recombination) [3, 4].  At times these rate terms must be included 
in the equation of continuity.  However, if sources are equal to sinks 
at all points or if these sources and sinks do not exist, the net charge 
flux or current density vector must satisfy the simplified (source and 
sink free) equation of continuity [1, 2]
\begin{equation} \label{e.2}
\frac{\partial \rho}{\partial t} + \nabla \cdot \textbf{J} = 0.
\end{equation}
A simple material follows Ohm's law given by [1, 2]
\begin{equation} \label{e.3}
\textbf{J} = \sigma \textbf{E}.
\end{equation}
For a simple material that follows Ohm's law, combining (2) and (3) gives
\begin{equation} \label{e.4}
\frac{\partial \rho}{\partial t} + \sigma \nabla \cdot \textbf{E} + \textbf{E}\cdot \nabla \sigma = 0.
\end{equation}
As noted above for a simple material $ \nabla \sigma = 0 $ and $ \nabla \epsilon = 0 $ so (1) and (4) yield
\begin{equation} \label{e.5}
\frac{\partial \rho}{\partial t} + \sigma \frac{\rho}{\epsilon}= 0.
\end{equation}
The traditional method to solve (5) is simply to state when charge 
density $ \rho $ is added that $ \sigma $ and $ \epsilon $ are still 
constants and define the electrical relaxation time of the conductor as
\begin{equation} \label{e.6}
\tau_{c}= \frac{\epsilon}{\sigma} 
\end{equation}
and integrate (5) from time \textit{t} = 0 when the 
charge density is $ \rho_{0} $ to general time \textit{t} obtaining
\begin{equation} \label{e.7}
\rho(t)=\rho_{0} \exp(-t/\tau_{c}).
\end{equation}
If a unipolar charge is placed in a material, the coulomb force will 
push the charges away from each other.  Eventually all charges will 
arrive at the surface so (7) indicates the net charge $ \rho $ at all 
points inside the material will decay to zero in approximately 5$ \tau_{c} $.  
The electrical conductivity of copper is 5.8 x 10$ ^{7} $ S/m and the 
dielectric constant $ \epsilon_{r} $ is 1 giving $ \tau_{c} $ = 
1.53 x 10$ ^{-19} $ second, whereas for fused quartz 
$ \sigma $ = 10$ ^{-17} $ S/m and  $ \epsilon_{r} $ = 5 giving 
$ \tau_{c} $ = 51.2 days; so, in many applications copper is considered a 
good conductor moving any injected charge to the surface of the 
conductor within 10$ ^{-18} $ second whereas quartz is considered an 
insulator suggesting that for the first several hours any charge 
introduced in quartz will behave as if it had remained essentially 
wherever it was placed [2].

In practice, it is found that metals and other good conductors 
follow (7) but that for insulator materials the initial decay appears 
hyperbolic with a decay rate dependent on the initial charge present.  As a 
result (7) is clearly not valid for both conductors and insulators.   
Later it will be argued that although (5) is the proper partial 
differential equation to describe the relaxation of the net charge with 
time, a more thorough investigation of Ohm's law shows that when charge 
is added to a material $ \sigma = \sigma(t) $ and, hence, the 
traditional method of solving 
(5) described above to obtain (7) is not actually valid. 
\medskip 
\begin{flushleft}
\textbf{Ohm's Law Revisited} \end{flushleft}
Today there is sufficient evidence to indicate matter is made up of 
\textit{neutral} 
atoms and groups of atoms called molecules.  Atoms are believed to be made up 
of a nucleus containing protons and neutrons with the nucleus surrounded 
by electrons.  Neutral atoms have  equal numbers of protons and electrons.  
An electron has a negative charge measured at -1.6 x 10$ ^{-19} $
 coulomb (C) and a proton has a positive charge of +1.6 x 10$ ^{-19} $ C.  
It is very difficult to remove a proton from an atom, but an electron 
can be more easily removed.  As a result, positive charges occur in 
matter when electrons are removed from neutral  atoms; and negative 
charges occur when electrons are added (attached) to neutral atoms or 
remain free.  This information and especially the fact that the intrinsic 
composition of a material is made up of \textit{neutral} atoms and 
\textit{neutral} molecules is 
important in understanding the nature of Ohm's law.  

The general charge flux equation for a single species of free charge 
is well known and given by ([3]; see their Eq. 18)
\begin{equation} \label{e.8}
\textbf{J}_{i} = \rho_{i}\textbf{v}_{d0}+\sigma_{i} \textbf{E}
-D_{i}\nabla \rho_{i}-G_{i}\rho_{i}\nabla T.
\end{equation}
Equation (8), which is also referred to as a constitutive 
law [5] can be obtained the traditional way by studying individually 
each transport process that can occur in a volume element at a point 
and then adding all these processes together [6] or by applying the 
contiguous-collision-averaging method to a point in space [7].  The nomenclature 
used here is that given in [7] in which the sign of charge $ s_{i} $ 
is explicit so that the charge density is [7]
\begin{equation} \label{e.9}
\rho_{i} = s_{i}q_{i}n_{i}
\end{equation}
where $ n_{i} $ is the number density of the charge species having a 
charge of magnitude $ q_{i} $ where $ q_{i} =Z_{i}q_{0}$ with 
$ q_{0}=+1.6 \times 10^{-19}$ coulomb being the fundamental unit of 
charge and $ Z_{i} $ being the number of fundamental units.  The term 
$ s_{i} $ is the sign of the charge where $ s_{i} $ = 1 for 
positive charge and $ s_{i} $ = -1 for negative charge.  As a result 
$ q_{i}n_{i} $ is always a positive quantity and the sign of $ \rho_{i} $
 is determined by the sign of $ s_{i} $. The introduction of the term 
$ s_{i} $ allows the mobility constant $ b_{i} $ to always be 
introduced as a positive quantity where the velocity of the charge is 
given by $ \textbf{v}_{i} = s_{i}b_{i}\textbf{E} $.  The conductivity of 
the i$ ^{th} $ species is [7]
\begin{equation} \label{e.10}
\sigma_{i} = s_{i}^{2}q_{i}n_{i}b_{i}
\end{equation}
The diffusion constant of the i$ ^{th} $ charged species is $ D_{i} $ 
and the thermophoresis coefficient is $ G_{i} $, but these need not be 
considered further here as this paper is restricted to homogeneous 
materials ($ \nabla \rho_{i}=0 $) at uniform temperature ($ \nabla T=0 $).  
The drift velocity $ \textbf{v}_{d0} $ in (8) of the material can often 
be important in liquids and gases and is zero for a stationary solid, 
liquid or gas.

Usually more than one species of charge exits in a material.  For example, 
in a neutral gas ionization removes electrons from neutral molecules 
producing free electrons as one species and positive ions as another 
species.  The electrons can then attach to neutral molecules and 
produce negative ions as a third species.  A similar result occurs 
in liquids due to dissociation.  In solids there can be many (conductor), 
some (semiconductor) or few (insulator) free electrons.  Electrons freed 
from atoms in solids also produce positive charged atoms, but these atoms 
cannot physically move.  However, by considering these charged atoms as 
holes, when an electron moves from its neutral atom to a previously 
charged atom (hole) the original hole disappears but a new hole is 
created.  This effective hole motion is  treated as a moving charged 
species [8].  In general the total charge flux is thus given by the 
sum of all the charge fluxes [7, 9]
\begin{equation} \label{e.11}
\textbf{J} =  \sum_{i} \textbf{J}_{i}.
\end{equation}
Consequently, (8) and (11) combine to define the total charge flux as
\begin{equation} \label{e.12}
\textbf{J} = \sum_{i} \rho_{i}\textbf{v}_{d0}+\sum_{i} \sigma_{i} \textbf{E}
-\sum_{i} D_{i}\nabla \rho_{i}-\sum_{i} G_{i}\rho_{i}\nabla T.
\end{equation}
In (12) the total or net volume charge density in the material at any 
point is given by [7]
\begin{equation} \label{e.13}
\rho = \sum_{i} \rho_{i}.
\end{equation}
Similarly the total electrical conductivity of the material at any 
point is [7]
\begin{equation} \label{e.14}
\sigma = \sum_{i} \sigma_{i}.
\end{equation}
With (13) and (14) it is possible to write (12) as [7]
\begin{equation} \label{e.15}
\textbf{J} = \rho\textbf{v}_{d0}+\sigma \textbf{E}
-\sum_{i} D_{i}\nabla \rho_{i}-(\sum_{i} G_{i}\rho_{i})\nabla T.
\end{equation}
As pointed out in [7] many materials have no convective 
drift $ \textbf{v}_{d0}=0 $, are homogeneous $ \nabla\rho_{i}=0 $ 
(i.e., are a simple material) and have a uniform temperature $ \nabla T=0 $.  
For these conditions (15) reduces to Ohm's law as given by (3).  Note 
that nothing in what has so far been presented requires the total 
electrical conductivity (14) to be a constant with time.  However, 
as Maxwell pointed out, Ohm's law would have little scientific 
value if a specific property of the material (i.e., the intrinsic 
conductivity) could not be defined [10].  This intrinsic electrical 
conductivity, which is not a function of time, will be clearly 
defined in the next section.
\medskip 
\begin{flushleft}
\textbf{Consequences of Ohm's Law} \end{flushleft}
In order to understand the concept of introducing free charge into a 
material the concept of intrinsic conductivity must first be understood.  
The intrinsic conductivity is the electrical conductivity of a material 
before the introduction of external free charges.  This can be best 
understood by examining the implications of (13) and (14) regarding 
a material's $ \rho $ and $ \sigma $ before external charge is 
introduced (Case 1).  Then the complication of introducing a single 
species of free charge (Case 2) can be easily followed.
\medskip 
\begin{flushleft}
\textbf{\space\space\space\space\space 
Case 1: (Charge neutrality; intrinsic conductivity)} \end{flushleft}
Because all atoms have equal amounts of positive and negative charge 
and because all materials are made up of atoms, all materials in their 
equilibrium or intrinsic state will be neutral.  Therefore, the 
intrinsic net charge must consist of equal numbers of positive and 
negative charge; consequently, the intrinsic net charge must be zero.  
This analysis can be easily extended to any number of \textit{sets} of different 
charge species but consider here the simplest situation were only one 
set of each charge (i.e., two charged species, one positive and the 
other negative) occurs.  When no other charges are present except 
those from the material's atoms and molecules, the equilibrium density 
is from (9) and (13)
\begin{equation} \label{e.16}
\rho(eq)\rightarrow \rho_{m}=\rho_{+}+\rho_{-}=
s_{+}q_{+}n_{+}+s_{-}q_{-}n_{-}=
s_{+}(q_{+}n_{+}-q_{-}n_{-})
\end{equation}
where it has been recalled that $ s_{+} =-s_{-}=1$.  From (10) 
and (14)
\begin{equation} \label{e.17}
\sigma(eq)\rightarrow \sigma_{m}=\sigma_{+}+\sigma_{-}=
s_{+}^{2}q_{+}n_{+}b_{+}+s_{-}^{2}q_{-}n_{-}b_{-}
\end{equation}
However, atomic charge neutrality requires the material to have no net 
charge ($ \rho_{m}=0 $ ) in   equilibrium  [8], so $ \rho_{m}=0 $ in 
(16) giving (for the special case $ n_{+}=n_{-} $ i.e., $ q_{+}=q_{-} $)
\begin{equation} \label{e.18}
q_{+}n_{+}=q_{-}n_{-} \equiv q_{m}n_{m}
\end{equation}
On the other hand, these free charges give rise to the conductivity 
of the material, so
\begin{equation} \label{e.19}
\sigma_{m}=s_{+}^{2}q_{m}n_{m}(b_{+}+b_{-})
\end{equation}
must be the intrinsic electrical conductivity of the material.  
This same procedure applies when $ n_{+} \neq n_{-} $ (i.e., some liquids, 
gases), but then $ q_{+} \neq q_{-} $, so (19) remains 
in the form of (17).  However, due to the atomic charge neutrality principle, 
in equilibrium $ \rho_{m}=0 $ will always be true.
\medskip 
\begin{flushleft}
\textbf{\space\space\space\space\space
Case 2: (Perturb with a single charge species $ \rho_{p} $)} \end{flushleft}
Next consider what happens when extrinsic charge $ \rho_{p} $ is uniformly 
placed in the material. For this case (9) and (13) require 
\begin{equation} \label{e.20}
\rho= \rho_{m}+\rho_{p}=\rho_{m}+s_{p}q_{p}n_{p}.
\end{equation}
If only interactions (collisions and transport) without reactions 
(changes in intrinsic ion concentrations) occur between the material 
and the inserted charge then $ \rho_{m} $ remains fixed. Hence, with the 
assumption of interaction without reaction $ \rho_{m}=0 $, so (20) becomes 
\begin{equation} \label{e.21}
\rho= \rho_{p}=s_{p}q_{p}n_{p}.
\end{equation}
Furthermore,  (10) and (14) require
\begin{equation} \label{e.22}
\sigma= \sigma_{m}+\sigma_{p}=\sigma_{m}+
s_{p}^{2}q_{p}n_{p}b_{p}=\sigma_{m}+s_{p}\rho_{p}b_{p}.
\end{equation}
Clearly, when the perturbation charge $ \rho_{p} $ is placed in the material, 
it is expected that coulomb repulsion will decay $ \rho_{p} $ away in 
time.  As a result, it is expected that $ \rho_{p}=\rho_{p}(t) $ which 
implies $ \sigma_{p}=\sigma_{p}(t) $, and, therefore, $ \sigma=\sigma(t) $ in 
(22).  The conclusion is clear: when charges are added to a material the 
electrical conductivity of the material is changed; and not until the charges
fully decay does the electrical conductivity of the material arive back at its 
intrinsic value.  As a result, $ \sigma $ in (5) is not independent of time; 
hence, (7) is not the general solution to (5).
\medskip 
\begin{flushleft}
\textbf{General Charge Decay Equation} \end{flushleft}
As discussed above, it is clear that (21) and (22) are required in (5); 
so, (5) becomes
\begin{equation} \label{e.23}
\frac{\partial \rho_{p}}{\partial t} + \frac{\sigma_{m}}{\epsilon}\rho_{p}
+\frac{s_{p}b_{p}}{\epsilon}\rho_{p}^{2}= 0.
\end{equation}
The material time constant can be defined with the intrinsic material 
conductivity as
\begin{equation} \label{e.24}
\tau_{m}=\frac{\epsilon}{\sigma_{m}}.
\end{equation}
The constants in the third term of (23) can be lumped together as
\begin{equation} \label{e.25}
\beta=\frac{s_{p}b_{p}}{\epsilon}.
\end{equation}
With the constants (24) and (25) the partial differential equation (5) 
{i.e.,  (23)} becomes
\begin{equation} \label{e.26}
\frac{\partial \rho_{p}}{\partial t} + \frac{\rho_{p}}{\tau_{m}}
+\beta\rho_{p}^{2}= 0.
\end{equation}
The solution to (26) is given by [11]
\begin{equation} \label{e.27}
\frac{\beta\tau_{m}\rho_{p}}{\beta\tau_{m}\rho_{p}+1}=
\frac{\beta\tau_{m}\rho_{p0}}{\beta\tau_{m}\rho_{p0}+1}\exp(-t/\tau_{m}).
\end{equation}
It is advantageous to define a time constant specified by the initial 
perturbation charge as
\begin{equation} \label{e.28}
\tau_{p}=(\beta\rho_{p0})^{-1}=\frac{\epsilon}{b_{p}(s_{p}\rho_{p0})}.
\end{equation}
With some rearranging (27) can now be written with the aid of (28) as
\begin{equation} \label{e.29}
\rho_{p}(t)=\frac{\rho_{p0}\exp(-t/\tau_{m})}
{1+\dfrac{\tau_{m}}{\tau_{p}}[1-\exp(-t/\tau_{m})]}
\end{equation}
which is the correct solution to (5) or (23).  Because (29) was derived 
without specifying the state (solid, liquid or gas) and the nature (good 
conductor, good insulator, etc.) of the material, (29) can be applied 
to all materials from very good conductors to very poor conductors 
(i.e., good insulators).
\medskip 
\begin{flushleft}
\textbf{Definition of a Good Conductor} \end{flushleft}
One way to define a good conductor is to require the third term in (26) 
to be negligible compared to the second term.  By this definition 
the criteria for a good conductor is $ \rho_{p}/\tau_{m}\gg\beta\rho_{p}^{2} $ 
or $ 1\gg\beta\tau_{m}\rho_{p} $.  However the largest value is at $ \rho_{p0} $,
so, the criteria for a good conductor must be $ 1\gg\beta\tau_{m}\rho_{p0} $.  
With the aid of (28) this criteria becomes $ \tau_{p}\gg\tau_{m} $.  
With this good conductor criterion (i.e., restriction) (29) reduces to
\begin{equation} \label{e.30}
\rho_{p}(t)\approx\rho_{p0}\exp(-t/\tau_{m}),  \textbf{\space\space\space 
only valid if }\tau_{p}\gg\tau_{m}.
\end{equation}
Equation (30) is similar to (7) except the material's intrinsic conductivity 
$ \sigma_{m} $ must be specified since, as shown in this paper, 
$ \sigma $ is not a constant. 
\medskip 
\begin{flushleft}
\textbf{Definition of a Good Insulator} \end{flushleft}
One way to define a good insulator is to require the second term in (26) 
to be negligible compared to the third term.  Hence, for a good 
insulator  $ \rho_{p}/\tau_{m}\ll\beta\rho_{p}^{2} $ 
or $ 1\ll\beta\tau_{m}\rho_{p} $ is required.  However, this can only 
be satisfied at the early stage of charge decay, because later, 
as $ t\rightarrow\infty $ so must $ \rho_{p} $ decay to zero, and, therefore, 
the requirement 
$ 1\ll\beta\tau_{m}\rho_{p} $ eventually always fails.  
At early time, the only time when the requirement can be met, 
the largest value is at $ \rho_{p0} $; 
so, the criteria for a good insulator is $ 1\ll\beta\tau_{m}\rho_{p0} $.  
With the aid of (28) this criterion for a good insulator 
becomes $ \tau_{p}\ll\tau_{m} $.  
\medskip 
\begin{flushleft}
\textbf{Comparison to Other "Insulator Laws"} \end{flushleft}
The neumerator and denominator of the single species charge decay 
equation (29) can be multiplied by $ \exp(t/\tau_{m}) $ to yield
\begin{equation} \label{e.31}
\rho_{p}(t)=\frac{\rho_{p0}}
{(1+\dfrac{\tau_{m}}{\tau_{p}})\exp(t/\tau_{m})-\dfrac{\tau_{m}}{\tau_{p}}}
\end{equation}
Equation (31) has a form somewhat similar to the so-called 
\textit{modified hyperbolic law} [12] as developed by Vellenga and 
Klinkenberg [13] 
in that both equations contain an exponential term and a 
time-constants-ratio term in their denominators.  However, the two 
equations are quite different in decay character in the 
range $ \tau_{m}< \sim 10\tau_{p} $.  However, (29), and 
hence also (31), reduces to the modified hyperbolic law for 
$ \tau_{m}\gg\tau_{p} $. 

  It is interesting to note that, in general, for $ t\ll\tau_{m} $ the 
exponential is given by $ \exp(t/\tau_{m})\approx1+t/\tau_{m} $ which 
when inserted into (31) gives
\begin{equation} \label{e.32}
\rho_{p}(t)
\approx\frac{\rho_{p0}}
{[1+\dfrac{t}{\tau_{p}}(1+\dfrac{\tau_{p}}{\tau_{m}})]}; 
\textbf{\space\space\space 
only valid if }t\ll\tau_{m}.
\end{equation}
For a very good insulator $ \tau_{p}\ll\tau_{m} $ and (32) reduces 
further to
\begin{equation} \label{e.33}
\rho_{p}(t)\approx\frac{\rho_{p0}}
{[1+\dfrac{t}{\tau_{p}}]}; 
\textbf{\space\space\space
only valid if }\tau_{p}\ll\tau_{m},
\textbf{ and only if }t\ll\tau_{m}.
\end{equation}
Although (33) is the \textit{hyperbolic law} of Bustin, 
et al. [14], the hyperbolic law is often (apparently improperly) assumed 
to be applicable over all time. As noted above, (33) shows the hyperbolic law 
is only valid for $ t\ll\tau_{m} $. 

Finally, note that $ \tau_{p}^{-1}=\beta\rho_{p0}=
[(s_{p}b_{p})/\epsilon]\rho_{p0} $; and, since $ \rho_{p0}=s_{p}q_{p}n_{p} $, 
the product in $ \tau_{p} $ always contains $ s_{p}^{2} $.  
Thus, $ \tau_{p} $ is always a positive quantity.  
\medskip 
\begin{flushleft}
\textbf{Discussion} \end{flushleft}
Equation (29) derived in this paper is a general equation for charge decay 
of unipolar ions which have an initial charge density $ \rho_{p0} $ and 
which have been uniformly introduced into any material: conductor 
or insulator; 
solid, liquid or gas. It is assumed that the material media into which 
the charge is introduced can be classified as a simple material; 
i.e., the material's electrical properties must be isotropic, linear and 
homogeneous.  In the derivation of (29) it is assumed that the inserted 
charge $ \rho_{p} $ only interacts with the material through collisions 
and transport and not through reactions which can change the intrinsic 
ion concentrations of the material.  In (29) the intrinsic material 
electrical relaxation time constant $ \tau_{m} $ is given by (24) 
where $ \epsilon $ is the material's permittivity and where $ \sigma_{m} $ 
is the intrinsic electrical conductivity of the material.  The time 
constant ratio in (29) is
\begin{equation} \label{e.34}
\frac{\tau_{m}}{\tau_{p}}=\frac{b_{p}s_{p}\rho_{p0}}{\sigma_{m}}
\end{equation}
where $ b_{p} $ (always a positive quantity) is the electrical mobility 
of the unipolar ions of initial volume charge density $ \rho_{p0} $ 
and $ s_{p} $ is the sign of the charge (either +1 or -1) of the 
unipolar ions.  The initial charge density of the unipolar ions in the material is
\begin{equation} \label{e.35}
\rho_{p0}=s_{p}q_{p}n_{p0}
\end{equation}
and depends on the number of ions per unit volume $ n_{p0} $, and the 
magnitude of the charge $ q_{p} $ of the ions where 
$ q_{p}=Z_{p}q_{p0} $ and where $ q_{0}=1.6\times10^{-19} $ C is the 
fundamental unit of charge and $ Z_{p} $ is the number of fundamental 
units.  Note that $ s_{p}\rho_{p0} $ is always positive.

The definition and distinction between a good conductor and a good 
insulator can be defined using the the two electrical relaxation 
times given in (24) and (28).  Namely, the criterion for a good 
conductor is $ \tau_{p}\gg\tau_{m} $ and for a good insulator 
is $ \tau_{p}\ll\tau_{m} $.  Since the mobilities are often of the 
same order of magnitude in a given material, these definitions also 
define a good conductor as $ n_{m}\gg n_{p} $ and a good insulator 
as $ n_{m}\ll n_{p} $.  These definitions and (29) are useful in 
specifying the type of charge decay (effectively exponential or 
early time hyperbolic).  Of course, the basic application definition 
of a good conductor still remains $ \tau_{m}\ll\tau_{use} $ 
 and for a good insulator  $ \tau_{m}\gg\tau_{use} $ where $ \tau_{use} $ is
 the use time for the process, investigation, operation, etc.
\medskip 
\begin{flushleft}
\textbf{Summary} \end{flushleft}
A general equation (29) has been developed which describes the time 
decay for a uniform distribution of unipolar charge inserted into any 
simple material (conductor or insulator; solid, liquid or gas) with an 
initial charge density $ \rho_{p0} $. The relaxation is controlled by 
both the intrinsic properties (conductivity $ \sigma_{m} $ and 
permittivity $ \epsilon $) of the material, and the extrinsic properties 
of the inserted charge, i.e., the initial volume charge 
density $ \rho_{p0} $ and the mobility $ b_{p} $ of the inserted charge.  
In (29) the charge density $ \rho_{p} $ is assumed not to chemically 
react with the material ($ \sigma_{m} $ invariant to $ \rho_{p} $) and 
decays in time $ t $ depending on the initial charge 
density $ \rho_{p0} $ defined by (35), on the material's intrinsic 
relaxation time constant $ \tau_{m} $ defined by (24), and on the 
perturbation time constant $ \tau_{p} $ defined by (28).
\medskip 
\begin{flushleft}
\textbf{References} 
\end{flushleft}

[1]Paul, C. R. and S. A Nasar, \textit{Introduction to Electromagnetic 
Fields}, 2nd Edition, McGraw-Hill: New York, (1987) pp. 111-114 and p. 234.

[2]Sadiku, M. N. O., \textit{Elements of Electromagnetics}, Holt, 
Rinehart and Winston: Orlando (1989), p. 193-196.

[3]Beuthe, T. G. and J. S. Chang, "Gas discharge phenomena" in 
\textit{Handbook of Electrostatic Processes}, Editors: J.-S. Chang, A. J. Kelly and J. M. 
Crowley; Marcel Dekker: New York (1995), pp. 147-193; 
see esp. p. 165, Eq. (17).

[4]Crowley, J. M. \textit{Fundamentals of Applied Electrostatics}, Wiley: 
New York (1986) Chapter 7.  especially pp. 115-120.

[5]Attens, P. and A. Castellanos, "Injection induced electrohydrodynamic 
flows" in \textit{Handbook of Electrostatic Processes}, Editors: J.-S. 
Chang, A. J. Kelly and J. M. Crowley; Marcel Dekker: New York (1995), 
pp. 121-146; see esp. pp. 124-125 and Eq. (10).

[6]Seaver, A. E., "Multicomponent transport equations in electrostatics," 
ESA Annual Meeting Proceedings, Laplacian Press: Morgan 
Hill, CA. (1995) pp. 193-209.

[7]Seaver, A. E., "Development of the charge flux equation using the 
contiguous collision averaging method," J. Electrostat., 
Vol. 46 (1999) pp. 177-191.

[8]Pierret, R. F., \textit{Semiconductor Fundamentals}, Second Edition, 
Addison-Wesley: New York (1988) For holes pp. 28-30 and for charge 
neutrality p. 47.

[9]Crowley, J. M., "Electrostatic fundamentals" in \textit{Handbook of 
Electrostatic Processes}, Editors: J.-S. Chang, A. J. Kelly and 
J. M. Crowley; Marcel Dekker: New York (1995) pp. 1-23; 
see esp. p. 9, Eq. (39).

[10]Maxwell, J. C. \textit{A Treatise on Electricity and Magnetism} 
Dover: New York,  3rd Edition, (1954) Vol. 1, pp. 362-363.

[11]Weast, R. C., Editor, \textit{CRC Handbook of Chemistry and Physics}, 
CRC Press: Boca Raton, FL 62nd Edition (1981-82) see Integrals 
p. A-37 Eq. 110.

[12]Pazda, R. J., and T. B. Jones and Y. Matsubara, "General theory for 
transient charge relaxation in a partially filled vessel," J. 
Electrostat., Vol. 32 (1994) pp. 215-231; see esp. p. 217, Eq. (3).

[13]Vellenga, S. J., and A. Klinkenberg, "On the rate of discharge of 
electrically charged hydrocarbon liquids," Chem. Eng. Sci., Vol. 
20 (1965) pp. 923-930.

[14]Bustin, W. M., I. Koszman and I. T. Tobye, "A new theory of static 
relaxation," Hydrocarbon Processing, Vol. 43 (1964) pp. 209-216.

\end{document}